\documentclass[twocolumn, tighten]{aastex62}
\usepackage{graphicx}
\usepackage{bm}

\received{November 2, 2017}
\revised{January 19, 2018}
\accepted{January 23, 2018}
\shorttitle{Enrichment of Zinc}
\shortauthors{Hirai et al.}

\begin{document}


\title{Enrichment of Zinc in galactic chemodynamical evolution models}

\correspondingauthor{Yutaka Hirai}
\email{yutaka.hirai@nao.ac.jp}

\author[0000-0002-5661-033X]{Yutaka Hirai}\altaffiliation{JSPS Research Fellow}
\affiliation{Department of Astronomy, Graduate School of Science, The University of Tokyo, 7-3-1 Hongo, Bunkyo-ku, Tokyo 113-0033, Japan}
\affiliation{Division of Theoretical Astronomy, National Astronomical Observatory of Japan, 2-21-1 Osawa, Mitaka, Tokyo 181-8588, Japan}
\author[0000-0001-8226-4592]{Takayuki R. Saitoh}
\affiliation{Earth-Life Science Institute, Tokyo Institute of Technology, 2-12-1 Ookayama, Meguro-ku, Tokyo 152-8551, Japan}
\author{Yuhri Ishimaru}\altaffiliation{Deceased November 18, 2017}
\affiliation{Department of Natural Sciences, College of Liberal Arts, International Christian University, 3-10-2 Osawa, Mitaka, Tokyo 181-8585, Japan}
\author{Shinya Wanajo}
\affiliation{Department of Engineering and Applied Sciences, Faculty of Science and Technology, Sophia University, 7-1 Kioicho, Chiyoda-ku, Tokyo 102-8554, Japan}
\affiliation{RIKEN, iTHES Research Group, 2-1 Hirosawa, Wako, Saitama 351-0198, Japan}

\begin{abstract}
The heaviest iron-peak element, Zn has been used as an important tracer of cosmic chemical evolution. Spectroscopic observations of the metal-poor stars in Local Group galaxies show that an increasing trend of [Zn/Fe] ratios toward lower metallicity. However, enrichment of Zn in galaxies is not well understood due to the poor knowledge of astrophysical sites of Zn as well as metal mixing in galaxies. Here we show {possible explanations for the observed trend} by taking into account electron-capture supernovae (ECSNe) as one of the sources of Zn in our chemodynamical simulations of dwarf galaxies. We find that the ejecta from ECSNe contribute to stars with [Zn/Fe] $\gtrsim$ 0.5. We also find that scatters of [Zn/Fe] in higher metallicity originate from the ejecta of type Ia supernovae. On the other hand, it appears difficult to explain the observed trends if we do not consider ECSNe as a source of Zn. These results come from inhomogeneous spatial metallicity distribution due to the inefficiency of metal mixing. We find that the optimal value of scaling factor for metal diffusion coefficient is $\sim$ 0.01 in the shear-based metal mixing model in smoothed particle hydrodynamics simulations. These results suggest that ECSNe can be one of the contributors to the enrichment of Zn in galaxies.
\end{abstract}

\keywords{methods: numerical  --- stars: abundances --- galaxies: abundances  --- galaxies: dwarf --- galaxies: evolution --- galaxies: formation}

\section{Introduction} \label{sec:intro}
Abundances of the heaviest iron-peak element, Zinc (Zn) in galaxies and interstellar medium (ISM) help us understand the stellar nucleosynthesis, cosmic chemical evolution, and metal mixing in galaxies.  Due to its volatile nature, Zn is not captured in dust grains. Measurements of Zn abundances in gas phase give us true gas-phase metallicity. In addition, Zn has been believed to be an ideal tracer of iron-peak elements because [Zn/Fe]\footnote{[A/B] = $\log_{10}({N_{\mathrm{A}}}/{N_{\mathrm{B}}})-\log_{10}({N_{\mathrm{A}}}/{N_{\mathrm{B}}})_{\odot}$, where $N_{\mathrm{A}}$ and $N_{\mathrm{B}}$ are the number of the elements A and B, respectively.} $\approx$ 0 for [Fe/H] $>$ $-$2 in the solar neighborhood. Therefore, gas phase abundances of Zn have been used as a tracer of metallicity in damped Lyman-$\alpha$ systems \citep[e.g.,][]{2005ARA&A..43..861W,2011A&A...530A..33V}.

\begin{figure}[htbp]
\epsscale{1.0}
\plotone{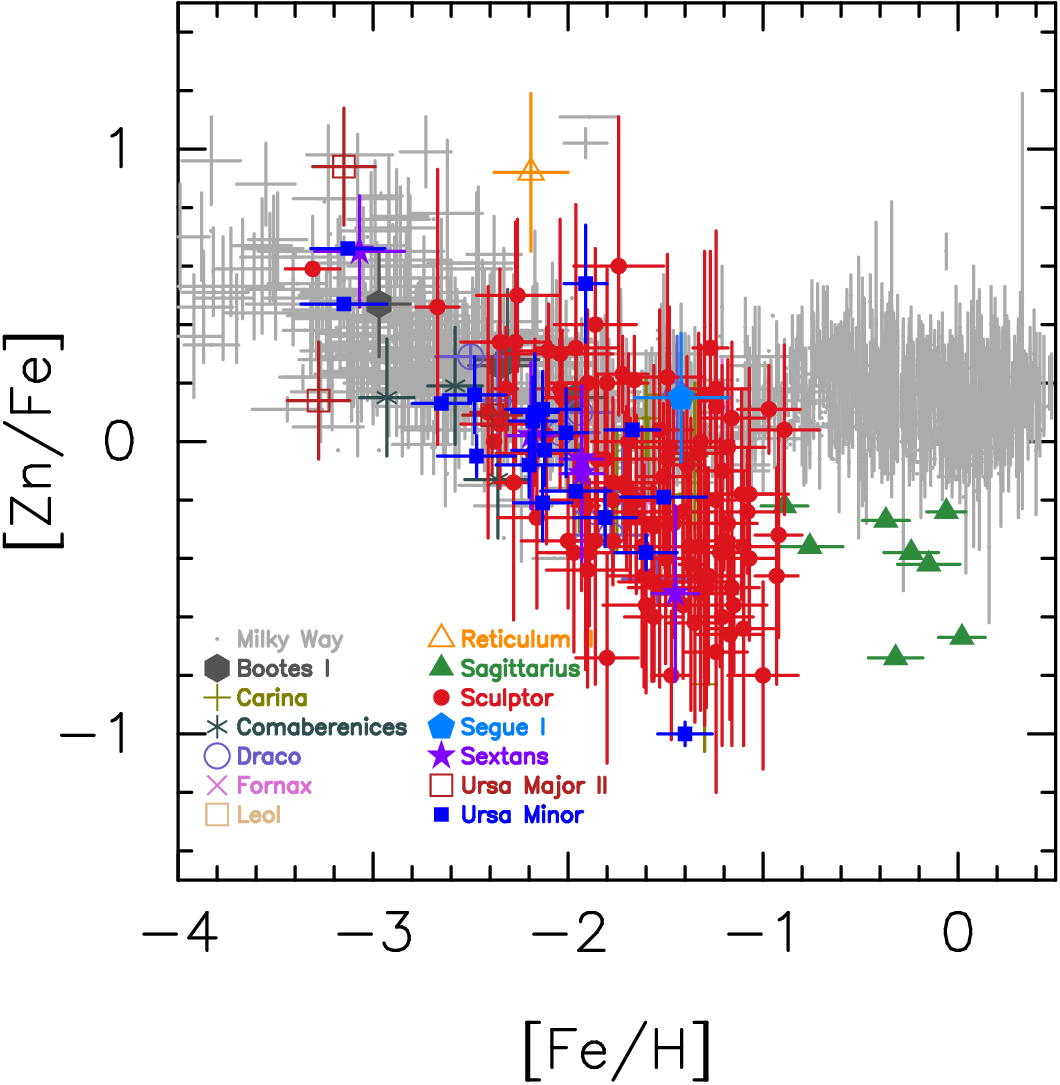}
\caption{[Zn/Fe] as a function of [Fe/H] for the Milky Way (light gray points) and Local Group dwarf galaxies (colored points). Bo\"otes I: \citet{2013ApJ...763...61G}, Carina: \citet{2003AJ....125..684S, 2012ApJ...751..102V}, Comaberenices: \citet{2010ApJ...708..560F}, Draco: \citet{2001ApJ...548..592S, 2009ApJ...701.1053C}, Fornax, Leo I: \citet{2003AJ....125..684S}, Reticulum II: \citet{2016ApJ...830...93J}, Sagittarius: \citet{2007A&A...465..815S}, Sculptor: \citet{2003AJ....125..684S, 2005AJ....129.1428G, 2015ApJ...802...93S, 2015A&A...574A.129S, 2017A&A...606A..71S}, Segue I: \citet{2014ApJ...786...74F}, Sextans: \citet{2001ApJ...548..592S, 2011PASJ...63S.523H}, Ursa Major II: \citet{2010ApJ...708..560F}, Ursa Minor: \citet{2001ApJ...548..592S, 2004PASJ...56.1041S, 2010ApJ...719..931C}. Error bars indicate the statistical as well as systematic errors given in each reference. All data are compiled using the SAGA database \citep{2008PASJ...60.1159S, 2011MNRAS.412..843S, 2017PASJ...69...76S, 2013MNRAS.436.1362Y}. \label{ZnFeobs}}
\end{figure}

High-dispersion spectroscopic observations have shown that [Zn/Fe] ratios increase toward lower metallicity at [Fe/H] $\lesssim$ $-$2 in the Milky Way \citep[e.g.,][]{2004A&A...416.1117C, 2004A&A...415..993N, 2007A&A...469..319N, 2009PASJ...61..549S,2017A&A...604A.128D} and the Local Group dwarf galaxies \citep[e.g.,][]{2010ApJ...708..560F, 2010ApJ...719..931C,2012ApJ...751..102V} as shown in Figure \ref{ZnFeobs}. {\citet{2017A&A...608A..46R} show that the slope of [Zn/Fe] as a function of [Fe/H] is $-$0.16 $\pm$ 0.05 for stars with $-$2.8 $\leq$ [Fe/H] $\leq$ $-$1.5. \citet{2009PASJ...61..549S} reported that scatters of [Zn/Fe] are $\sim$ 0.6--0.7 dex at [Fe/H] $\leq$ $-$2.0 in the Milky Way halo. At [Fe/H] $>$ $-$2.0, these scatters decrease to $\sim$ 0.5--0.6 dex and flat [Zn/Fe] ratios can be seen in the Milky Way halo.} \citet{2017A&A...606A..71S} find that there are large star-to-star scatters, $-$0.8 $\lesssim$ [Zn/Fe] $\lesssim$ 0.4, and a decreasing trend of [Zn/Fe] ratios at [Fe/H] $\gtrsim$ $-$1.8 in the Sculptor dwarf spheroidal galaxy (dSph). These observational features should be associated with astrophysical sites of Zn.

Astrophysical sites of Zn are highly complicated. Since massive stars have relatively shorter lifetimes than those of lower-mass stars, Zn ejected by supernovae (SNe) can be a dominant source of the enrichment of Zn at low metallicity ([Fe/H] $\lesssim$ $-$2). Stars more massive than $\sim$ 10 $M_{\sun}$ explode as core-collapse supernovae (CCSNe). They synthesize $^{64}$Zn during complete Si-burning and $^{66-70}$Zn by neutron-capture process \citep{1995ApJS..101..181W, 2006ApJ...653.1145K}. Several chemical evolution studies have been conducted to understand the Galactic enrichment history of Zn \citep{1993A&A...272..421M, 1995ApJS...98..617T, 2000A&A...359..191G, 2004A&A...421..613F, 2006isna.confE..37I, 2006ApJ...653.1145K}. \citet{1995ApJS...98..617T} suggest that the amount of Zn in the observation can be explained if they reduce the Fe yields by a factor of two. \citet{2006ApJ...653.1145K} show that the CCSN yields of \citet{1997NuPhA.616...79N} cannot give enough Zn to explain the observation. 

Hypernovae (HNe) have been suggested as a possible astrophysical site of Zn \citep{2002ApJ...565..385U,2005ApJ...619..427U,2006ApJ...653.1145K,2007ApJ...660..516T}. HNe produce $\sim$ 1 dex larger kinetic energies than those of normal CCSNe \citep[e.g.,][]{2009ApJ...692.1131T} and are thought to be observed as long gamma-ray bursts \citep[GRBs, ][]{2004ApJ...607L..17P, 2007ApJ...657L..73G}. The outward Si-burning regions in HNe synthesize larger amounts of Zn. \citet{2006ApJ...653.1145K} show that if half of the stars heavier than 20 $M_{\sun}$ explode as HNe, the [Zn/Fe] ratio increases $\sim$ 1 dex compared to the prediction of \citet{1997NuPhA.616...79N}. \citet{2007ApJ...660..516T} suggest that the increasing trend can be reproduced if EMP stars reflect each yield of HN with a different progenitor mass. 

Electron capture SNe (ECSNe) can also be the astrophysical sites of Zn. The lowest-mass ($\lesssim$ 10 $M_{\sun}$) progenitors of CCSNe that develop oxygen-neon-magnesium cores cause this type of SNe \citep[e.g.,][]{1980PASJ...32..303M, 1982Natur.299..803N, 1984A&A...133..175H, 1984ApJ...277..791N, 1987ApJ...322..206N, 1987ApJ...318..307M}. The explosion occurs in stars with core mass of 1.367 $M_{\sun}$ \citep{2013ApJ...771...28T} when the electrons captured by $^{24}$Mg and $^{20}$Ne remove the pressure support. Hydrodynamic simulations show that the explosion energy of ECSN is $\sim$ 10$^{50}$ erg \citep[e.g.,][]{2006A&A...450..345K, 2008A&A...485..199J, 2012PTEP.2012aA309J, 2011ApJ...726L..15W}. \citet{2015MNRAS.446.2599D} predicted that the mass range of progenitors of ECSNe is 9.8--9.9 $M_{\sun}$ at solar metallicity by their stellar evolution calculation. This range depends on the treatment of mass-loss rate, the efficiency of third dredge-up, and convection \citep[e.g.,][]{2007A&A...476..893S, 2008ApJ...675..614P, 2013ApJ...772..150J, 2014ApJ...797...83J, 2015ApJ...810...34W, 2017PASA...34...56D}. \citet{2009ApJ...695..208W, 2011ApJ...726L..15W} estimated that the upper limit of the fraction of ECSNe in all CCSNe was about 30 \% based on their nucleosynthesis calculations. ECSNe can be observed as optically bright SNe characterized by the short plateau with a faint tail luminosity curve \citep[e.g.,][]{2013ApJ...771L..12T, 2013MNRAS.434..102S, 2014A&A...569A..57M}. The crab nebula (SN 1054) is one of the most promising candidates of an ECSN remnant \citep[e.g.,][]{1982ApJ...253..696D, 1982Natur.299..803N}.

Nucleosynthesis studies based on two-dimensional hydrodynamic simulations of an 8.8 $M_{\sun}$ ECSN show that  ECSNe produce all the stable isotopes of Zn in neutron-rich ejecta with the electron fraction (proton-to-nucleon ratio) of $\sim$ 0.4 -- 0.5 \citep{2011ApJ...726L..15W, 2018ApJ...852...40W}. ECSNe also synthesize a small amount of Fe. These features of nucleosynthesis lead to higher values of [Zn/Fe] than those of normal CCSNe. In addition to Zn, ECSNe may contribute to the enrichment of light trans-iron elements \citep[e.g.,][]{2011ApJ...726L..15W, 2014A&A...568A..47H, 2017ApJ...837....8A}, $^{48}$Ca \citep{2013ApJ...767L..26W}, and $^{60}$Fe \citep{2013ApJ...774L...6W}. There are several studies on the enrichment of $r$-process elements by ECSNe \citep[e.g.,][]{1999ApJ...511L..33I, 2004ApJ...600L..47I}. However, their role in the enrichment of Zn in galaxies has not yet been studied.

Ejecta from SNe mix into the surrounding ISM. Newly formed stars inherit the abundances of the mixed gas in the star-forming region. We thus need to properly treat the metal mixing in galaxies to constrain the astrophysical sites of Zn. \citet{2012MNRAS.425..969P} show that the fractions of extremely metal-poor (EMP) stars will be overestimated if they do not consider metal diffusion. The metal-mixing process also affects the carbon abundances of metal-poor stars \citep{2017ApJ...834...23S}. \citet{2017ApJ...838L..23H} suggest that timescale of metal mixing is $\lesssim$ 40 Myr to explain the observed abundances of $r$-process elements in dSphs.

Dwarf galaxies are the ideal objects to study to gain an understanding of complex chemical enrichment histories because of their simple structures \citep[e.g.,][]{2012A&A...538A..82R, 2013ApJ...774..103L, 2015ApJ...807..154B}. \citet{2015ApJ...814...41H, 2017MNRAS.466.2474H} clarify the enrichment histories of $r$-process elements in dwarf galaxies by their chemodynamical simulations. \citet{2016A&A...588A..21R} show that it is necessary to introduce the metal-mixing scheme in particle-based simulations to reproduce the observed low scatters of $\alpha$-element abundances in metal-poor stars. Recent observations of metal-poor stars in the Local Group dwarf galaxies enable us to compare with such simulation results \citep[e.g.,][]{2010ApJ...708..560F, 2010ApJ...719..931C,2012ApJ...751..102V, 2017A&A...606A..71S}.

The main purpose of this study is to clarify the enrichment of Zn in dwarf galaxies using a series of high-resolution galactic chemodynamical simulations. We aim to constrain the astrophysical sites of Zn as well as the metal-mixing processes in galaxies by comparing our simulations with observations of metal-poor stars. In Section \ref{sec:method}, we describe our models adopted in this study. In Section \ref{sec:results}, we show the star-formation histories (SFHs), the metallicity distributions, and the $\alpha$-element abundances computed by our fiducial model. In Section \ref{EnrichmentofZn}, we show the enrichment of Zn computed in our dwarf galaxy models. We discuss the astrophysical sites of Zn as well as the efficiency of metal mixing in galaxies implied from the enrichment histories of Zn. In Section \ref{Conclusions}, we show our conclusions.
\section{Method} \label{sec:method}
\subsection{Code}
\subsubsection{$N$-body/SPH code, \textsc{asura}}
Here we describe the $N$-body/smoothed particle hydrodynamics (SPH) code, \textsc{asura} \citep{2008PASJ...60..667S,2009PASJ...61..481S} adopted in this study. Details of the code are described in \citet{2008PASJ...60..667S, 2015ApJ...814...41H, 2017MNRAS.466.2474H}. Gravity is calculated using the Tree method \citep{1986Natur.324..446B}. We adopt a density-independent formulation of SPH (DISPH) to compute hydrodynamics \citep{2013ApJ...768...44S}. This scheme enables us to treat fluid instability in a contact discontinuity properly \citep{2016ApJ...823..144S}. We implement a Fully Asynchronous Split Time-integrator for a self-gravitating fluid (FAST) algorithm to reduce the computation cost \citep{2010PASJ...62..301S}. We also implement a time-step limiter to treat strong shock regions correctly \citep{2009ApJ...697L..99S}.

We use the metallicity-dependent cooling/heating functions from 10 to 10$^9$ K generated by \textsc{cloudy} \citep{1998PASP..110..761F, 2013RMxAA..49..137F}. We adopt the ultraviolet background heating table of \citet{2012ApJ...746..125H}. The effect of hydrogen self-shielding is implemented following \citet{2013MNRAS.430.2427R}. 

We probabilistically select gas particles to become star particles when they satisfy the three conditions: (1) $\nabla\cdot\bm{v}<$ 0 ($\bm{v}$ is the velocity of gas), (2) high number density ($>$ 100 cm$^{-3}$), and (3) low temperature ($<$ 1000 K). Each star particle is treated as a single stellar population (SSP). We adopt the initial mass function (IMF) of \citet{2001MNRAS.322..231K} from 0.1 to 100 $M_{\sun}$. The number of Ly$\alpha$ photons in a H$_{\rm II}$ region is computed using \textsc{p\'egase} \citep{1997A&A...326..950F}.

We adopt Chemical Evolution Library (\textsc{celib}) to compute stellar feedback and chemical evolution \citep{2016ascl.soft12016S, 2017AJ....153...85S}. We assume that stars more massive than 8 $M_{\sun}$ explode as CCSNe and distribute the thermal energy of $\sim$ 10$^{51}$ erg to surrounding gas particles. We also assume that a certain fraction ($f_{\rm HN}$) of stars more massive than 20 $M_{\sun}$ explode as HNe. They distribute the thermal energy of $\sim$ 10$^{52}$ erg to surrounding gas particles. We adopt $f_{\rm HN}$ = 0, 0.05, and 0.5 in this study. The value of $f_{\rm HN}$ = 0.05 is taken from the observed rate of long GRBs \citep{2004ApJ...607L..17P, 2007ApJ...657L..73G}. The value of $f_{\rm HN}$ = 0.5 is taken from the chemical evolution model of \citet{2006ApJ...653.1145K}. We adopt the nucleosynthesis yields of \citet[hereafter, N13]{2013ARA&A..51..457N} in our fiducial models. We also test the nucleosynthesis yields of \citet[hereafter, CL04]{2004ApJ...608..405C} at $Z\geq10^{-5}Z_{\sun}$ and \citet{2012ApJS..199...38L} at $Z<10^{-5}Z_{\sun}$ for comparison. Type Ia supernovae (SNe Ia) are assumed to occur in stars with 3--6 $M_{\sun}$ in this simulation. We adopt the power law delay time distribution with the index of $-1$ according to \citet{2012PASA...29..447M}. We set the minimum delay time as 10$^8$ yr \citep{2008PASJ...60.1327T}. We adopt the model N100 of \citet{2013MNRAS.429.1156S} for the nucleosynthesis yields of SNe Ia. {We assume that each SN Ia produces 0.74 $M_{\sun}$ of Fe and does not produce Zn.}

\subsubsection{ECSN model}
We newly consider the role of ECSNe in our simulation. We adopt the nucleosynthesis yields of model e8.8 in \citet{2018ApJ...852...40W}; the same as those in \citet{2011ApJ...726L..15W, 2013ApJ...767L..26W, 2013ApJ...774L...6W}. When an ECSN occurs, the thermal energy of 9 $\times$ 10$^{49}$ erg is distributed to the surrounding gas particles according to \citet{2018ApJ...852...40W}. The mass range of ECSN progenitors is taken from the stellar evolution calculations of \citet{2015MNRAS.446.2599D}. At $Z$ = 0.0001, the lower and upper progenitor masses of ECSNe are 8.2 $M_{\sun}$ and 8.4 $M_{\sun}$, respectively. At $Z$ = 0.02, the mass range becomes 9.8 -- 9.9 $M_{\sun}$. We also adopt the mass range computed by \citet{2007PhDT.......212P}. Their models show more pronounced metallicity dependences on the mass ranges than those of \citet{2015MNRAS.446.2599D}. Also, we consider the case of a constant mass range of 8.5 -- 9.0 $M_{\sun}$. {Table \ref{ECSNe} summarizes the mass ranges of ECSN progenitors adopted in this study.} We discuss the effects of adopted mass ranges in Section \ref{site}.

\begin{deluxetable}{lcccccc}[htbp]
   \tabletypesize{\scriptsize}
   \tablecaption{The mass ranges of ECSN progenitors. \label{ECSNe}} 
   \tablecolumns{2}
   \tablewidth{0pt}
   \tablehead{
      \colhead{}&
      \multicolumn{2}{l}{\citet{2015MNRAS.446.2599D}}&
      \multicolumn{2}{l}{\citet{2007PhDT.......212P}}&
      \multicolumn{2}{l}{Constant mass range}\\
      \colhead{$Z$}&
      \colhead{$M_{\rm l}$}&
      \colhead{$M_{\rm u}$}&
      \colhead{$M_{\rm l}$}&
      \colhead{$M_{\rm u}$}&
      \colhead{$M_{\rm l}$}&
      \colhead{$M_{\rm u}$}\\
      \colhead{}&\colhead{($M_{\sun}$)}&\colhead{($M_{\sun}$)}&\colhead{($M_{\sun}$)}&\colhead{($M_{\sun}$)}&\colhead{($M_{\sun}$)}&\colhead{($M_{\sun}$)}}
      \startdata
      0.02&9.8&9.9&9.0&9.3&8.5&9.0\\
      0.008&9.5&9.6&8.7&9.3&8.5&9.0\\
      0.004&8.8&9.0&8.4&9.1&8.5&9.0\\
      0.001&8.3&8.4&7.6&8.4&8.5&9.0\\
      0.0001&8.2&8.4&6.9&8.2&8.5&9.0\\
      0.00001&8.2&8.4&6.4&8.2&8.5&9.0\\
      \enddata
     \tablecomments{From left to right, columns show metallicity ($Z$), the lower ($M_{\rm l}$) and upper ($M_{\rm u}$) masses of ECSN progenitors.}
\end{deluxetable}

\subsection{Nucleosynthesis yields of Zinc}\label{Nucleosynthesis}
We adopt the nucleosynthesis yields of Zn for ECSNe \citep{2018ApJ...852...40W}, normal CCSNe (N13 or CL04), and HNe (N13). {The former (ECSNe) and latters (CCSNe and HNe) are based on the solar-metallicity and metallicity-dependent models, respectively. Table \ref{YieldTable} lists the yields of  ECSNe, CCSNe, and HNe for selected metallicities adopted in this study. The solar-metallicity model of an ECSN is used for all the other metallicities because of its insensitivity to the initial compositions \citep{2011ApJ...726L..15W, 2018ApJ...852...40W}.} \citet{2018ApJ...852...40W} show that all stable Zn isotopes are predominantly synthesized in neutron-rich ejecta with the electron fraction from 0.4 to 0.5. ECSNe produce sufficient amounts of such neutron-rich ejecta. This leads to the production of Zn in ECSNe. On the other hand, higher entropy ejecta from HNe than those from normal CCSNe result in a strong $\alpha$-rich freeze out from nuclear statistical equilibrium. This leads to greater production of $^{64}$Zn in HNe than that in normal CCSNe (e.g., N13). 

\begin{deluxetable*}{lllllllllll}[htbp]
   \tablecaption{Yields of Fe and Zn of each type of SN. \label{YieldTable}} 
   \tablecolumns{2}
   \tablewidth{0pt}
   \tablehead{
      \colhead{}&
      ECSNe&
      \multicolumn{5}{l}{CCSNe}&
      \multicolumn{4}{l}{HNe}\\
      \colhead{Elements}&
      e8.8&
      15&
      20&
      25&
      30&
      40&
      20&      
      25&
      30&      
      40\\
            &
      ($M_{\sun}$)&
      ($M_{\sun}$)&
      ($M_{\sun}$)&
      ($M_{\sun}$)&
      ($M_{\sun}$)&
      ($M_{\sun}$)&
      ($M_{\sun}$)&      
      ($M_{\sun}$)&
      ($M_{\sun}$)&      
      ($M_{\sun}$)
      }
      \startdata
     $Z$ = 0&&&&&&&&&&\\
Fe&
3.07$\times$10$^{-3}$&
7.24$\times$10$^{-2}$&
7.23$\times$10$^{-2}$&
7.38$\times$10$^{-2}$&
7.46$\times$10$^{-2}$&
8.00$\times$10$^{-2}$&
8.49$\times$10$^{-2}$&
9.94$\times$10$^{-2}$&
1.64$\times$10$^{-1}$&
2.63$\times$10$^{-1}$\\
Zn&
1.13$\times$10$^{-3}$&
1.23$\times$10$^{-4}$&
8.34$\times$10$^{-5}$&
2.56$\times$10$^{-6}$&
3.08$\times$10$^{-10}$&
4.44$\times$10$^{-11}$&
3.85$\times$10$^{-4}$&
2.66$\times$10$^{-4}$&
5.90$\times$10$^{-4}$&
6.96$\times$10$^{-4}$\\
     $Z$ = 0.001&&&&&&&&&&\\
Fe&
3.07$\times$10$^{-3}$&
7.20$\times$10$^{-2}$&
7.23$\times$10$^{-2}$&
7.18$\times$10$^{-2}$&
7.26$\times$10$^{-2}$&
7.93$\times$10$^{-2}$&
8.19$\times$10$^{-2}$&
1.52$\times$10$^{-1}$&
2.04$\times$10$^{-1}$&
2.64$\times$10$^{-1}$\\
Zn&
1.13$\times$10$^{-3}$&
6.84$\times$10$^{-5}$&
3.32$\times$10$^{-5}$&
4.86$\times$10$^{-5}$&
8.21$\times$10$^{-5}$&
1.08$\times$10$^{-4}$&
3.59$\times$10$^{-4}$&
5.66$\times$10$^{-4}$&
8.54$\times$10$^{-4}$&
7.28$\times$10$^{-4}$\\
     $Z$ = 0.004&&&&&&&&&&\\
Fe&
3.07$\times$10$^{-3}$&
7.05$\times$10$^{-2}$&
7.08$\times$10$^{-2}$&
6.88$\times$10$^{-2}$&
7.31$\times$10$^{-2}$&
7.38$\times$10$^{-2}$&
2.63$\times$10$^{-2}$&
7.81$\times$10$^{-2}$&
1.50$\times$10$^{-1}$&
2.74$\times$10$^{-1}$\\
Zn&
1.13$\times$10$^{-3}$&
1.18$\times$10$^{-4}$&
8.47$\times$10$^{-5}$&
2.18$\times$10$^{-4}$&
2.64$\times$10$^{-4}$&
5.71$\times$10$^{-4}$&
6.03$\times$10$^{-5}$&
4.03$\times$10$^{-4}$&
5.44$\times$10$^{-4}$&
1.00$\times$10$^{-3}$\\
     $Z$ = 0.02&&&&&&&&&&\\
Fe&
3.07$\times$10$^{-3}$&
6.55$\times$10$^{-2}$&
6.15$\times$10$^{-2}$&
5.85$\times$10$^{-2}$&
6.09$\times$10$^{-2}$&
5.22$\times$10$^{-2}$&
9.29$\times$10$^{-3}$&
9.35$\times$10$^{-2}$&
7.31$\times$10$^{-2}$&
2.57$\times$10$^{-1}$\\
Zn&
1.13$\times$10$^{-3}$&
7.13$\times$10$^{-5}$&
2.25$\times$10$^{-4}$&
9.16$\times$10$^{-4}$&
7.08$\times$10$^{-6}$&
3.02$\times$10$^{-3}$&
1.60$\times$10$^{-4}$&
1.12$\times$10$^{-3}$&
1.32$\times$10$^{-4}$&
3.33$\times$10$^{-3}$\\
      \enddata
     \tablecomments{The first column shows metallicity or names of elements. From the second left to right, columns show yields of Fe and Zn. Yields of ECSN are taken from \citet{2018ApJ...852...40W}. We take the solar-metallicity model of ECSNe because they are insensitive to the initial composition. For yields of CCSNe and HNe, we take those from N13.}
\end{deluxetable*}

\begin{figure}[htbp]
\epsscale{1.0}
\plotone{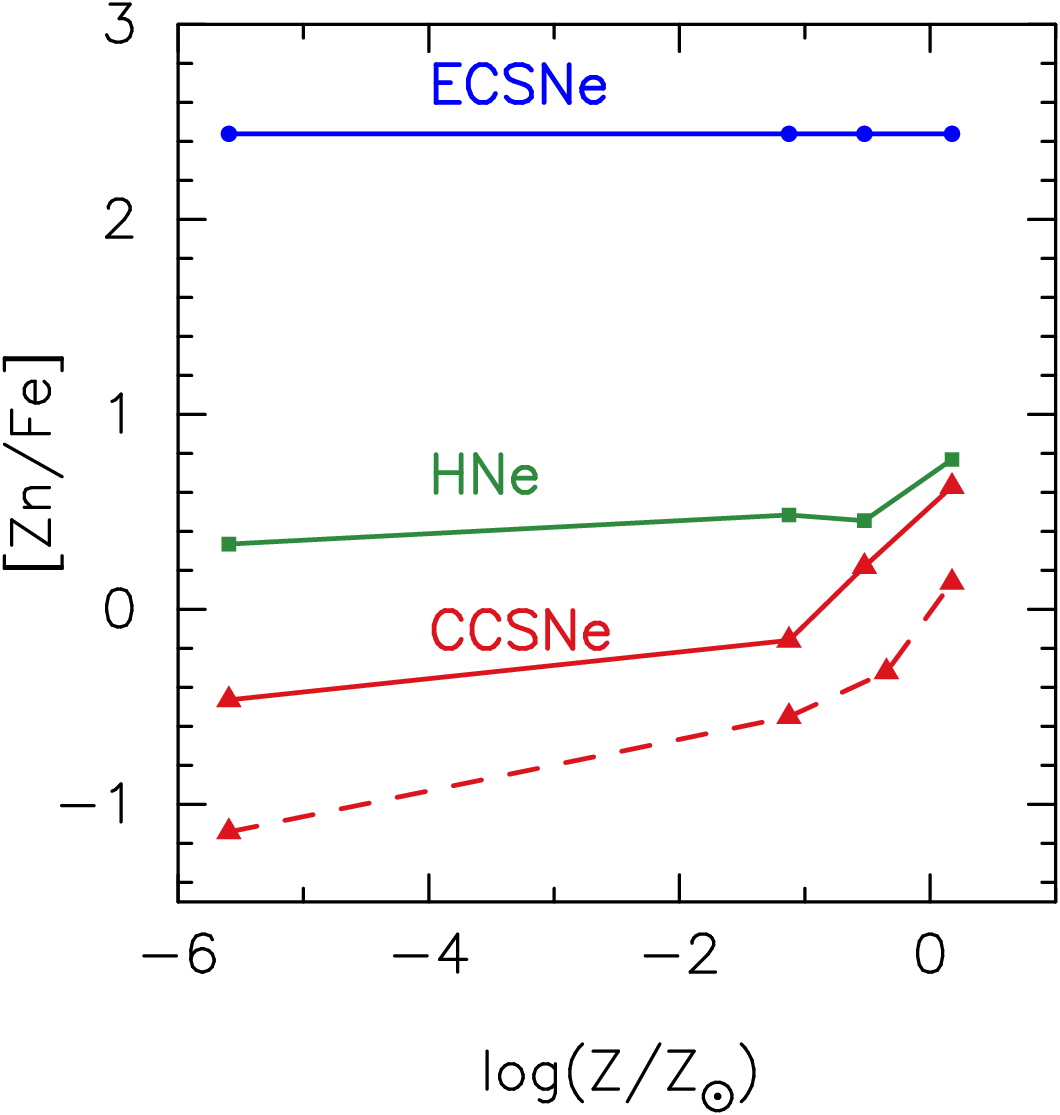}
\caption{The IMF integrated yields of [Zn/Fe] as a function of metallicity. Blue circles, green squares, and red triangles connected with solid lines represent the IMF weighted yields of ECSNe, HNe, and CCSNe, respectively. The ECSN yields are taken from \citet{2018ApJ...852...40W}. We do not consider the metallicity dependence of the yields of ECSNe. The HN and CCSN yields are taken from N13. The red dashed line shows the CCSN yields of CL04. Yields for $Z$ = 0 are plotted at $\log_{\rm 10} (Z/Z_{\sun})$ = $-$5.6. We adopt $Z_{\sun}$ = 0.0134 \citep{2009ARA&A..47..481A}.\label{Yields}}
\end{figure}

Figure \ref{Yields} shows the IMF-integrated yields of [Zn/Fe] as a function of metallicity. ECSNe produce higher [Zn/Fe] ratios than those of CCSNe and HNe. This is because ECSNe synthesize sufficiently small amounts of Fe compared to the other types of SNe. HNe synthesize sufficient Fe as well as Zn. This feature of nucleosynthesis results in the lower values of [Zn/Fe] than those of ECSNe. The yield of Zn from a CCSN is sensitive to the location of the mass cut that divides the ejected material and the remnant core. We discuss the effects of different yields of SNe in Appendix \ref{sec:yields}.

\begin{figure}[htbp]
\epsscale{1.0}
\plotone{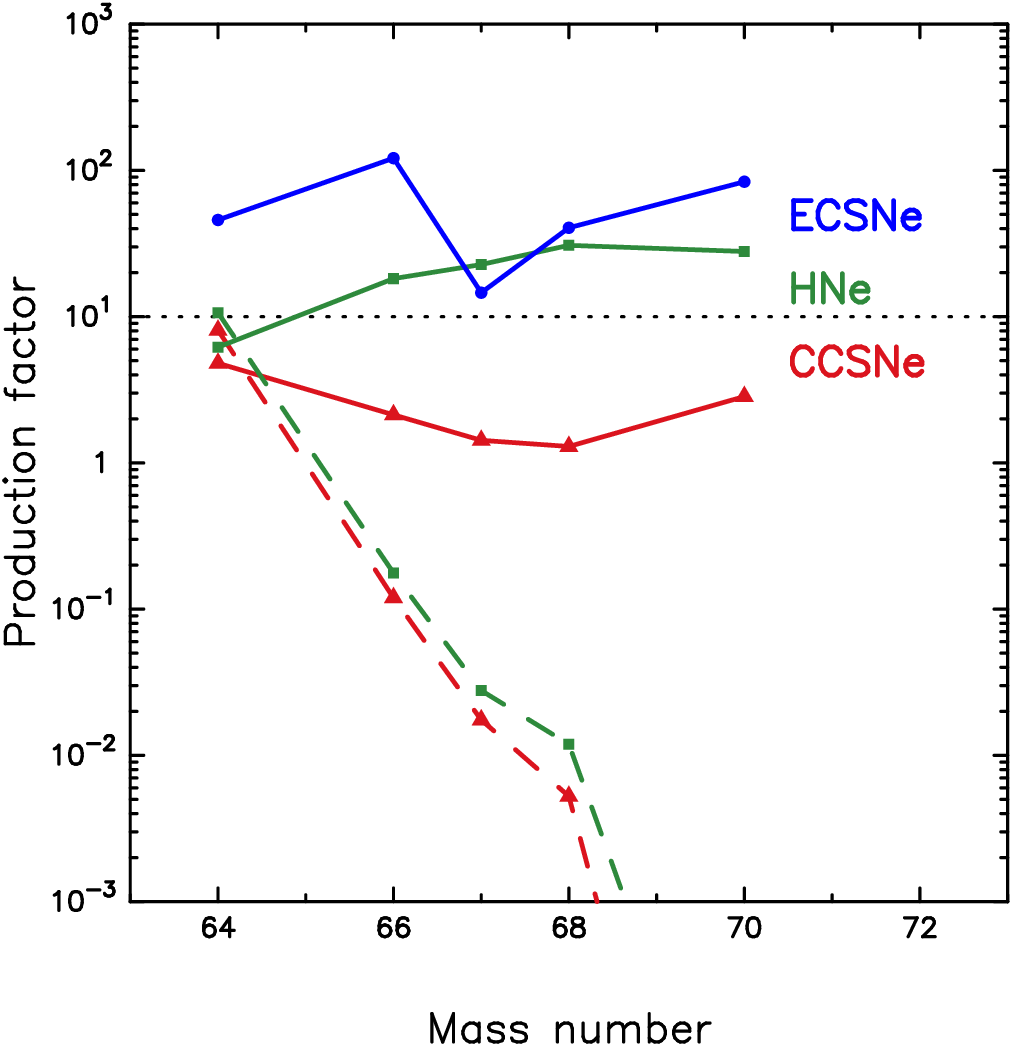}
\caption{{Production factors of Zn isotopes. Blue circles, green squares, and red triangles connected with solid lines represent the production factors of ECSNe (8.8 $M_{\sun}$), HNe (25 $M_{\sun}$), and CCSNe (15 $M_{\sun}$) at the solar metallicity, respectively. Green and red dashed lines denote the production factors of HNe (25 $M_{\sun}$) and CCSNe (15 $M_{\sun}$) at $Z$ = 0, respectively. Yields of isotopes are taken from \citep{2018ApJ...852...40W} for ECSNe and N13 for HNe and CCSNe. The black dotted line indicates the production factor of 10 that is taken as the lower bound for each astrophysical site to be a main contributor of a given isotope \citep[e.g.,][]{2007PhR...442..269W}.} \label{Isotopes}}
\end{figure}

{Figure \ref{Isotopes} shows the mass fractions of Zn isotopes relative to the solar values (production factors). The black dotted line indicates the production factor of 10 that is taken as the lower bound for each astrophysical site to be a main contributor of a given isotope \citep[e.g.,][]{2007PhR...442..269W}. At $Z$ = 0, the production factors of $^{66, 67, 68, 70}$Zn from HNe and CCSNe are significantly lower than this lower bound. The production factors of these isotopes increase with metallicity. This is because the weak $s$-process, which synthesizes $^{66, 67, 68, 70}$Zn in CCSNe, is more efficient at higher metallicity \citep{2006ApJ...653.1145K}. \citet{2011MNRAS.414.3231K} show that the ratio of $^{64}$Zn/$^{66, 67, 68, 70}$Zn continuously decreases toward higher metallicity. However, they find that the ratio is too low to explain the solar isotopic ratios at $Z$ = $Z_{\sun}$. On the other hand, ECSNe synthesize all isotopes of Zn (independent of metallicities) in a nuclear statistical equilibrium \citep{2011ApJ...726L..15W, 2018ApJ...852...40W}. As shown in Figure \ref{Isotopes}, the production factors of the Zn isotopes in ECSNe are between $1.5~\times~10$ and $1.2~\times~10^2$. This suggests that a contribution of ECSNe with a rate of several percent of all CCSNe can explain most of the solar isotopic abundances of Zn.}

\subsection{Metal-mixing model}
In SPH simulations without a metal-mixing scheme, the metals inherited from SNe are locked in a gas particle throughout the galaxy evolution. In this study, we take a shear-based metal-mixing model \citep{2010MNRAS.407.1581S, 2017AJ....153...85S}. The time derivative of $i$th metal ($Z_{i}$) follows diffusion equation, 
\begin{eqnarray}
\frac{{\rm{d}}Z_i}{{\rm{d}}t} &=&  \nabla(D{\nabla}Z_i),\nonumber\\
D &=& C_{\rm{d}}|S_{ij}|h^{2},
\end{eqnarray}
where $D$ is the diffusion coefficient, $C_{\rm d}$ is the scaling factor for diffusion coefficient, $S_{ij}$ is the trace-free shear tensor, and $h$ is the smoothing length of SPH particle. \citet{2017ApJ...838L..23H} show that the value of $C_{\rm{d}}$ $\gtrsim$ 0.01 is appropriate to explain the observational trends of $r$-process elements in dSphs. We discuss the efficiency of metal mixing in Section \ref{sec:mixing}.

\subsection{Isolated dwarf galaxy model}
We adopt an isolated dwarf galaxy model such as those in \citet{2009A&A...501..189R, 2012A&A...538A..82R, 2015ApJ...814...41H, 2017MNRAS.466.2474H, 2017ApJ...838L..23H}. We take the same structural parameters in \citet{2017ApJ...838L..23H}. The initial total number of particles and the gravitational softening length are 2$^{18}$ and 7.8 pc, respectively. We show that our main results do not strongly depend on the resolution of simulations (Appendix \ref{sec:resolution}). The total mass of halo is 7 $\times$ 10$^8$ $M_{\sun}$. The final stellar mass of our model is 5 $\times$ 10$^6$ $M_{\sun}$, which is similar to those of the Sculptor and Leo I dSphs \citep{2012AJ....144....4M}. We show the resulting SFHs, metallicity distribution, and Mg abundances in the next section. Table \ref{models} lists all the models adopted in this study.

\begin{deluxetable}{llll}[htbp]
   \tabletypesize{\scriptsize}
   \tablecaption{List of models. \label{models}} 
   \tablecolumns{2}
   \tablewidth{0pt}
   \tablehead{
      \colhead{Model}&
      \colhead{$C_{\rm d}$}&
      \colhead{ECSN mass range}&
      \colhead{$f_{\rm HN}$}}
      \startdata
      A&0.01&\citet{2015MNRAS.446.2599D}&0.05\\  
      B&0.01&\citet{2007PhDT.......212P}&0.05\\ 
      C&0.01&8.5 -- 9.0 $M_{\sun}$&0.05\\ 
      D&0.01&\nodata&0.5\\  
      E&0.01&\nodata&0.05\\ 
      F&0.01&\citet{2015MNRAS.446.2599D}&\nodata\\  
      G&0.01&8.5 -- 9.0 $M_{\sun}$&\nodata\\ 
      H&0.001&\citet{2015MNRAS.446.2599D}&0.05\\  
      I&0.1&\citet{2015MNRAS.446.2599D}&0.05\\ 
      \enddata
     \tablecomments{From left to right, columns show the names of models, the scaling factors for metal diffusion, the mass ranges of ECSN progenitors, and the HN fractions.}
\end{deluxetable}

\section{Chemodynamical evolution of dwarf galaxies} \label{sec:results}
\begin{figure}[htbp]
\epsscale{1.0}
\plotone{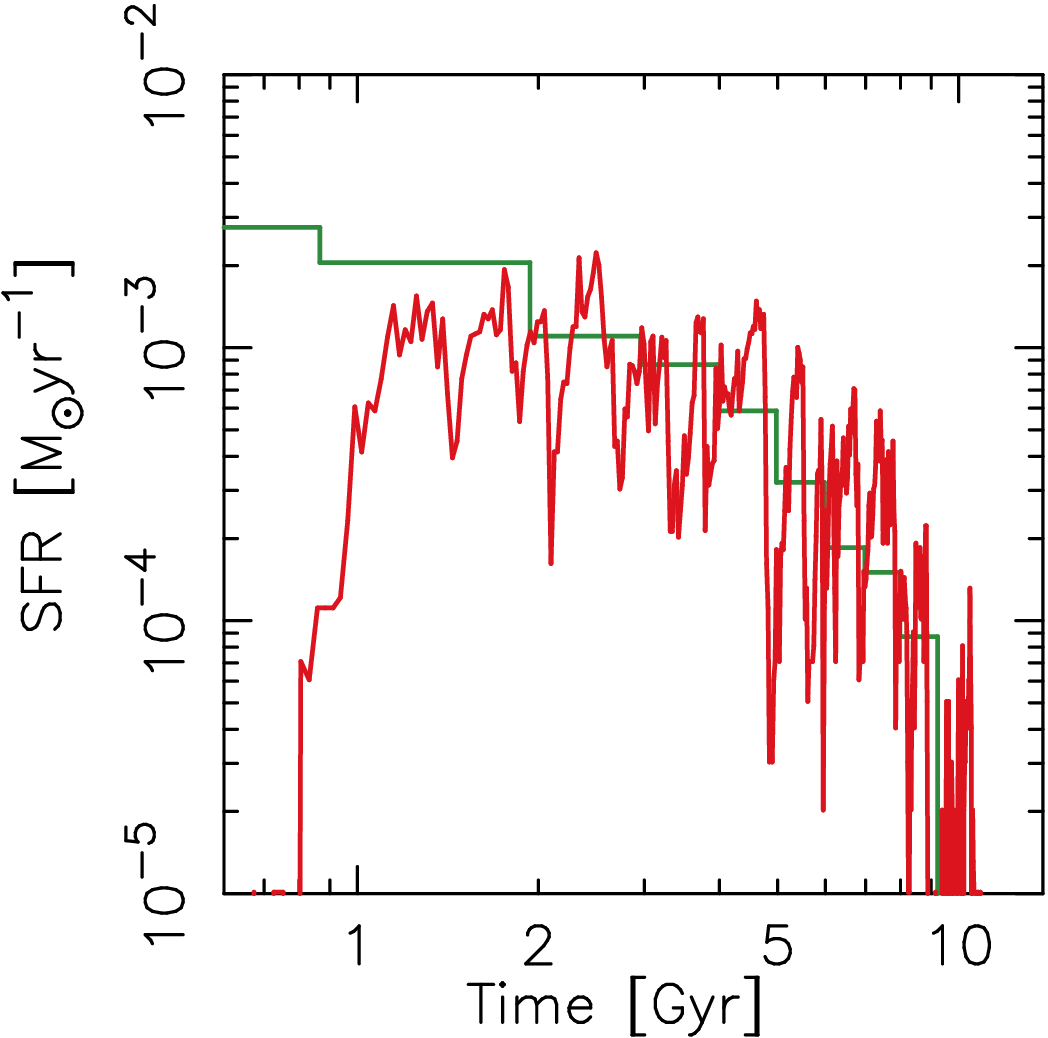}
\caption{Time variations of SFRs in model A (red curve). {The green line shows the SFH of the Sculptor dSph estimated from the color-magnitude diagram \citep{2012A&A...539A.103D}.} \label{SFR}}
\end{figure}

Here we show the SFHs, the metallicity-distribution functions, and the $\alpha$-element abundances computed in our models. Although we do not intend to construct models that are relevant to specific Local Group dwarf galaxies, we confirm that our models have properties similar to those of the observed dwarf galaxies, as follows:

Figure \ref{SFR} shows the time variations of star-formation rates (SFRs) in model A. Star formation begins after a sufficient gas fall onto the central region of the galaxy. The oscillating behavior of SFRs is due to discontinuous SN feedbacks. As shown in this figure, a typical SFR in this model is $\sim$ 10$^{-3}$ $M_{\sun}$yr$^{-1}$. This value is roughly consistent with the SFHs of Local Group dSphs such as Sculptor and Fornax estimated from color-magnitude diagrams \citep{2012A&A...544A..73D, 2012A&A...539A.103D}.

\begin{figure}[htbp]
\epsscale{1.0}
\plotone{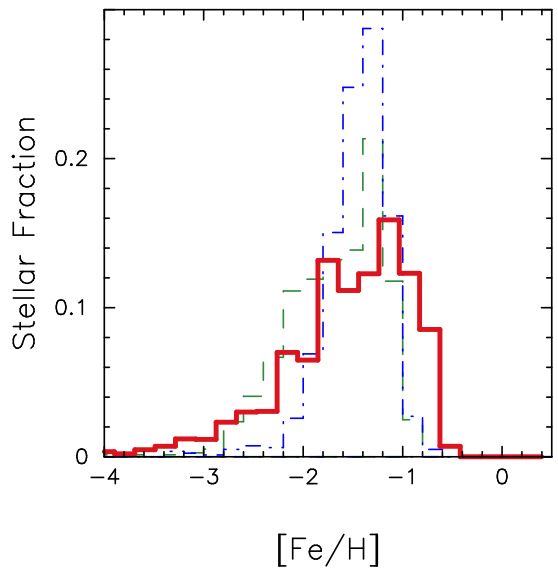}
\caption{Predicted and observed metallicity distribution functions. The red histogram represents the metallicity distribution function of model A at 13.8 Gyr from the beginning of the simulation. The green dashed, and blue dash-dotted histograms show the observed metallicity distribution functions of the Sculptor and Leo I dSphs \citep{2009ApJ...705..328K,2010ApJS..191..352K,2012AJ....144..168K}. All observed data are compiled using the SAGA database \citep{2017PASJ...69...76S}.\label{MD}}
\end{figure}

Figure \ref{MD} shows the metallicity distribution function of model A at 13.8 Gyr from the beginning of the simulation. The median metallicity and the final stellar mass are [Fe/H] = $-$1.34 and $M_{*}$ = 3.72 $\times$ 10$^6$ $M_{\sun}$, respectively. These values are consistent with those of Local Group dwarf galaxies such as the Sculptor dSph ([Fe/H] = $-$1.68 and $M_{*}$ = 3.9 $\times$ 10$^6$ $M_{\sun}$) and the Leo I dSph ([Fe/H] = $-$1.45 and $M_{*}$ = 4.9 $\times$ 10$^6$ $M_{\sun}$) within the observational errors \citep{2013ApJ...779..102K}. 

\begin{figure}[htbp]
\epsscale{1.0}
\plotone{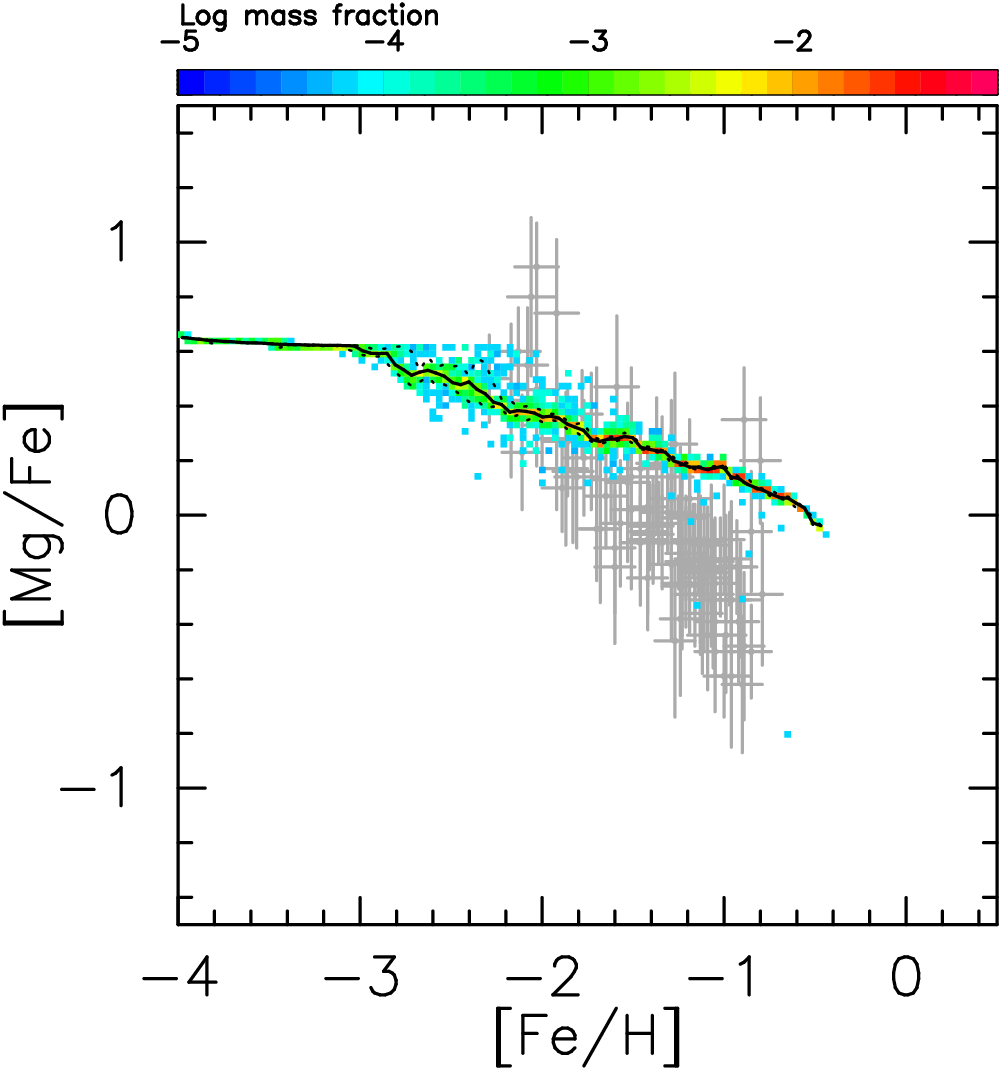}
\caption{[Mg/Fe] as a function of [Fe/H] in model A. The color-coded stellar mass fractions are displayed in the logarithmic scale. {The solid black curve shows the median value of computed data at each [Fe/H] bin. The dotted curves show the 5 \% and 95 \% significance levels. Gray dots represent the observed values for Sculptor dSph \citep{2010ApJS..191..352K}. We plot observed values with internal errors of $\Delta$[Fe/H] $<$ 0.15, $\Delta$[Mg/Fe] $<$ 0.30 to be consistent with high resolution data (Hill et al. in prep) following \citet{2012ASPC..458..297H}.} \label{MgFe}}
\end{figure}

Figure \ref{MgFe} shows the $\alpha$-element abundance ratios ([Mg/Fe]) as a function of [Fe/H] in model A. Stars with [Fe/H] $\lesssim$ $-$2.8 in this model show star-to-star scatters of [Mg/Fe] less than 0.1 dex. Such small scatters of [Mg/Fe] also are reported for the EMP stars in some Local Group galaxies \citep[e.g., ][]{2015ARA&A..53..631F}. The observed values of [Mg/Fe] decrease as the metallicity increases for stars with [Fe/H] $\gtrsim$ $-$2.5. This trend can be interpreted as a consequence of the contribution of SNe Ia that do not produce $\alpha$-elements. In our model, the values of [Mg/Fe] start to decrease at [Fe/H] $\approx$ $-$2.5, which is consistent with the case of the Sculptor dSph reported in \citet{2017PASJ...69...76S}.

The slope of computed [Mg/Fe] as a function of [Fe/H] is shallower than the observed one. This may be due to the different SFH between our model and Sculptor or modeling of SNe Ia. \citet{2012A&A...539A.103D} show that Sculptor has steadily decreasing SFH from 14 to 7 Gyr ago. On the other hand, our models have a constant SFH over 9 Gyr (Figure \ref{SFR}). When the SFH is peaked at the very early stages, the computed slope of [Mg/Fe] tends to be steeper than the slopes of models with SFH peaked at later phases \citep{2015ApJ...799..230H}. Another possibility to cause the different slope can be attributed to our modeling of SNe Ia. We model the delay time distribution of SNe Ia with power-law distribution with the minimum delay time of 0.1 Gyr. \citet{2009ApJ...707.1466K} show that SNe Ia rates should be very low in [Fe/H] $\lesssim$ $-$1 to reproduce observed $\alpha$-element abundances. \citet{2015ApJ...799..230H} suggest that the minimum delay time of SNe Ia is estimated to be 0.5 Gyr from the $\alpha$-element abundances in dSphs. Since discussing these effects is beyond the scope of this paper, we do not discuss these possibilities further here.

\section{Enrichment of Zinc}\label{EnrichmentofZn}
\subsection{Enrichment of Zinc at [Fe/H] $\lesssim$ $-$2.5}
\begin{figure}[htbp]
\epsscale{1.0}
\plotone{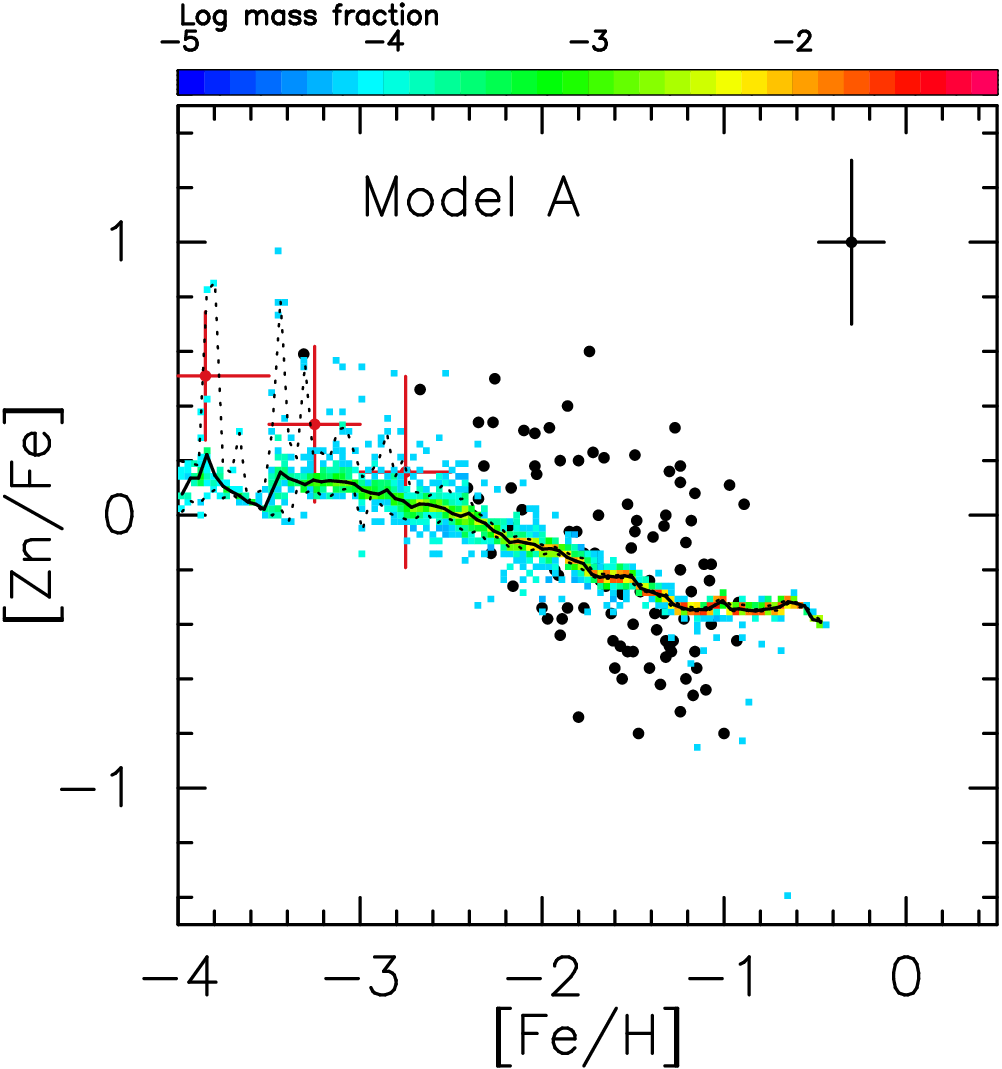}
\caption{Same as Figure \ref{MgFe}, but for [Zn/Fe] as a function of [Fe/H] in model A. Black dots denote observed values of Sculptor dSph \citep{2003AJ....125..684S, 2005AJ....129.1428G, 2015ApJ...802...93S, 2015A&A...574A.129S, 2017A&A...606A..71S}. {Typical error bars of the observation \citep{2017A&A...606A..71S} are shown in the top right corner of the figure.} The red points show the average values of [Zn/Fe] in the Milky Way halo stars \citep{2009PASJ...61..549S}. The vertical error bars on red points indicate the difference between the maximum and minimum values of [Zn/Fe] in each metallicity bin. The horizontal bars on red points represent the range of [Fe/H] in each bin.\label{ZnFe}}
\end{figure}

The observation of [Zn/Fe] is characterized by an increasing trend toward lower metallicity. In Figure \ref{ZnFe}, we plot observed data of Sculptor and the Milky Way halo. We plot the binned Milky Way halo data at [Fe/H] $<$ $-$2.5 because observed data of dSphs are not enough to compare with our models. \citet{2017A&A...606A..71S} show that the abundance ratios of [Zn/Fe] in Sculptor are consistent with those of the Milky Way halo at [Fe/H] $\lesssim$ $-$2. Here we mainly focus on figuring out conditions to form stars with [Zn/Fe] $\gtrsim$ 0.5, which are seen in the observations. We also compare the slopes of [Zn/Fe] as a function of [Fe/H] in our models and observations.

Figure \ref{ZnFe} shows [Zn/Fe] as a function of [Fe/H] computed in model A. Stars with [Zn/Fe] $\gtrsim$ 0.5 at [Fe/H] $\lesssim$ $-$2 reflect the high [Zn/Fe] ratios in the ejecta of ECSNe (Figure \ref{Yields}). On the other hand, the ejecta from HNe {and CCSNe contribute to increasing the average values of [Zn/Fe]. The average values of [Zn/Fe] are determined by the IMF weighted values of Zn yield of all astrophysical sites of Zn. In model A, the average values of Zn are lower than those of observations at [Fe/H] $\lesssim$ $-$3. In Section \ref{site}, we show that these values are related to the rates of ECSNe and HNe.}

The stellar [Zn/Fe] ratios of model A slightly increase toward lower metallicity. We performed the chi-squared linear fitting to our data at $-$4.0 $<$ [Fe/H] $<$ $-$2.5. We choose this metallicity range because the effects of SNe Ia are negligible at [Fe/H] $<$ $-$2.5 and there are not enough observed data at [Fe/H] $<$ $-$4.0. The linear fitting of our result shows that the slope of [Zn/Fe] as a function of [Fe/H] is $-$0.12 $\pm$ 0.01. On the other hand, the linear fitting of the observed data of the Milky Way halo \citep[SAGA database,][]{2008PASJ...60.1159S, 2011MNRAS.412..843S, 2013MNRAS.436.1362Y} shows the slope of $-$0.26 $\pm$ 0.04. We take the observed data of the Milky Way halo because the number of observed data of dwarf galaxies is not enough to analyze in this metallicity range. {The flatter slope than that of the observations is due to the adopted mass ranges of the ECSN progenitors in low metallicity. We adopt the stellar evolution model of \citet{2015MNRAS.446.2599D} in model A. They show that the mass range of ECSN varies only 0.1 $M_{\sun}$ from $Z$ = $10^{-5}$ to $10^{-3}$ $Z_{\sun}$.  This almost constant mass range in low metallicity makes it difficult to reproduce the increasing trend toward lower metallicity. At [Fe/H] $\gtrsim$ $-$2.9, SNe Ia start contributing to decrease the [Zn/Fe] ratio toward higher metallicity.}

\begin{figure}[htbp]
\epsscale{1.0}
\plotone{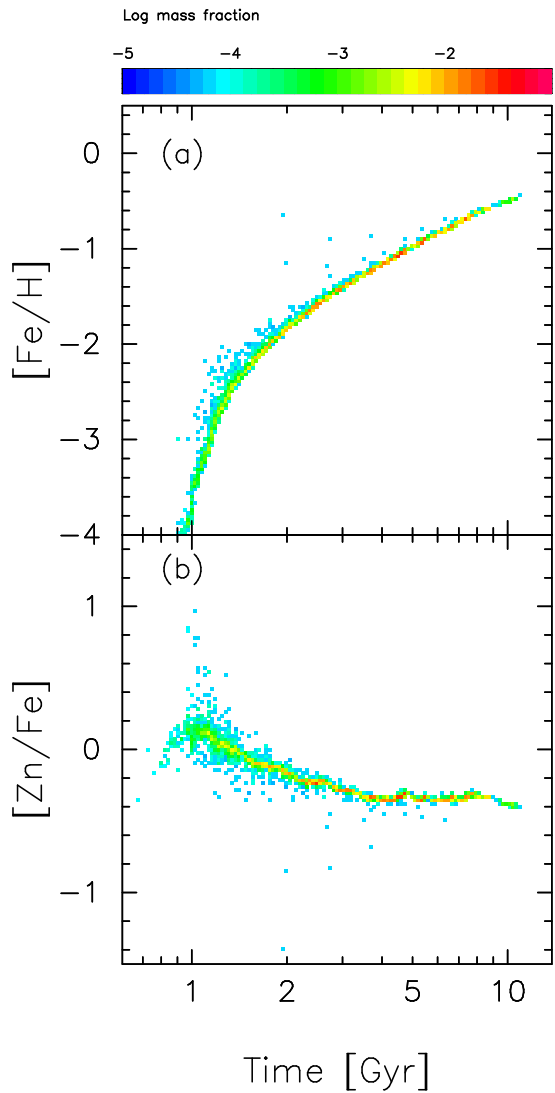}
\caption{(a) [Fe/H] and (b) [Zn/Fe] as a function of time in model A. The color-coded stellar mass fractions are displayed in the logarithmic scale. \label{AgeFe}}
\end{figure}

{The stars of model A with the highest [Zn/Fe] ratios for given metallicity bins have an increasing trend toward lower metallicity. Figure \ref{ZnFe} shows that the highest [Zn/Fe] ratio at [Fe/H] $=$ $-$2.5 is [Zn/Fe] = 0.4. The highest value increases to [Zn/Fe] = 1.0 at [Fe/H] $=$ $-$3.5. This trend} comes from the inhomogeneity of the spatial metallicity distribution in the early epoch of galaxy evolution. Figure \ref{AgeFe} shows the time evolutions of [Fe/H] and [Zn/Fe]. As shown in Figure \ref{AgeFe}a, stars with [Fe/H] $<$ $-$2 are formed until 2 Gyr from the beginning of the simulation. ECSNe synthesize an appreciable amount of Zn with a small amount of Fe (Figure \ref{Yields}). When the first ECSN occurs in the galaxy, stars subsequently formed around the ejecta of ECSNe have high values of [Zn/Fe] ($\sim$ 1). These stars can only be formed in the early epoch of galaxy evolution ($t$ $<$ 2 Gyr in this model, Figure \ref{AgeFe}b).  The fraction of ECSNe out of all SNe is 3.1 \% at $Z$ = 0.0001 in this model. Due to the low event rate of ECSNe, the Zn abundances in the ISM are highly inhomogeneous during the early epoch of galaxy evolution. As time passes, an increase of Fe by SNe as well as metal mixing reduces the fraction of gases with high [Zn/Fe] values.

Figure \ref{ZnFemass} shows [Zn/Fe] versus [Fe/H] for the models B and C (Table \ref{models}) assuming different mass ranges of ECSN progenitors as well as HNe with $f_{\rm HN}$ = 0.05. As shown in this figure, both models tend to have stars with higher [Zn/Fe] ratios at lower metallicities than those in model A (Figure \ref{ZnFe}) as a consequence of adopting wider ranges of progenitors of ECSNe than the stellar evolution models in \citet{2015MNRAS.446.2599D}.  The fractions of ECSNe out of all CCSNe are 38.2 \% and 6.9 \% in models B and C, respectively (3.1 \% in model A).  Although the slope of [Zn/Fe] is still shallower than that of the observation, the average values of [Zn/Fe] are consistent with the observation in model C. These results suggest that the increasing trends of [Zn/Fe] toward lower metallicity can be reproduced if the rates of ECSNe are higher in lower metallicity.

The rates of ECSNe in low metallicity are determined by the mass range and lifetimes of progenitors of ECSNe. However, we cannot constrain the mass range of ECSNe in this model because there are large uncertainties in models of galaxies as well as of stellar evolution. {Stripped-envelope stars in close binaries can be the progenitors of ECSNe \citep{2013ApJ...778L..23T, 2015MNRAS.451.2123T, 2016MNRAS.461.2155M}. The mass range for ECSNe in binary systems is predicted to be wider than that for single stars \citep{2004ApJ...612.1044P, 2017ApJ...850..197P}. \citet{2017MNRAS.471.4275Y} suggest that ultra-stripped SNe synthesize large amounts of iron-peak elements. \citet{2018ApJ...852...40W} show that the SNe from the low-mass end of the progenitors with iron-cores can also synthesize Zn as large as that of ECSNe.  In the case of chemodynamical simulations of galaxies, we currently treat star particles as SSPs, i.e., the yields produced by star particles are IMF weighted values of CCSNe, HNe, or ECSNe. Future chemodynamical simulations of galaxies that can resolve each star will make it possible to discuss the effects of different yields of individual SNe.

\begin{figure}[htbp]
\epsscale{1.0}
\plotone{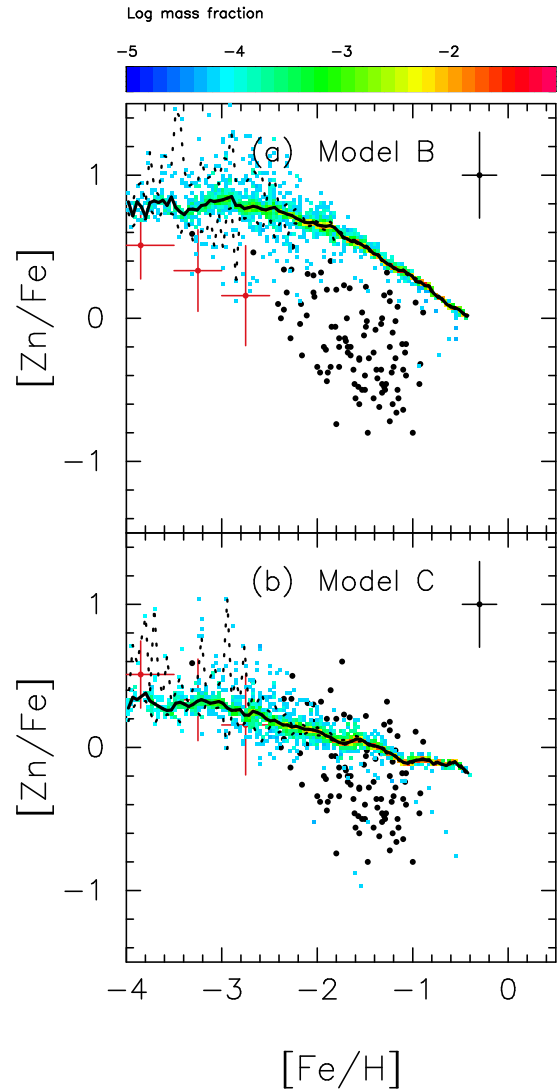}
\caption{Same as Figure \ref{ZnFe} but for (a) model B and (b) model C. Model B adopts the metallicity dependent mass range of ECSN progenitors computed by the stellar evolution model in \citet{2007PhDT.......212P}. Model C assumes that stars from 8.5 to 9.0 $M_{\sun}$ become ECSNe in all the metallicity range. \label{ZnFemass}}
\end{figure}

\subsection{Enrichment of Zinc at [Fe/H] $\gtrsim$ $-$2.5}
At higher metallicity, the Milky Way halo, and dSphs have different trends of [Zn/Fe] as a function of [Fe/H] (Figure \ref{ZnFeobs}). Sculptor has increasing trends toward lower metallicity while the Milky Way halo stars have constant [Zn/Fe] ratios. \citet{2017A&A...606A..71S} imply that there are star-to-star scatters of [Zn/Fe] in Sculptor at [Fe/H] $\gtrsim$ $-$1.8. They show that several stars with the same metallicity have apparently different [Zn/Fe] ratios. However, they cannot confirm these scatters of [Zn/Fe] ratios because of their low signal-to-noise data. The scatters are mostly consistent with their error-bars.

For $-$2.5 $\lesssim$ [Fe/H] $\lesssim$ $-$1.0, {the increasing contribution of} SNe Ia makes the slope of [Zn/Fe] steeper than that in lower metallicity. Since SNe Ia do not produce Zn, stars formed from the gas polluted by the ejecta of SNe Ia have low [Zn/Fe] values. Scatters of [Zn/Fe] values for [Fe/H] $\gtrsim$ $-$2 reflect the inhomogeneity of [Zn/Fe] abundances affected by SNe Ia. {Stars with [Zn/Fe] $<$ $-$0.5 are formed under substantial influence from the ejecta of SNe Ia. As shown in Figure \ref{AgeFe}b, all stars with [Zn/Fe] $<$ $-$0.5 are formed within 4 Gyr from the beginning of the simulation.

Scatters of [Zn/Fe] at [Fe/H] $\gtrsim$ $-$2.5 are not different from those of $\alpha$-elements in our models. Figure \ref{Sigma} shows standard deviations of [Mg/Fe] and [Zn/Fe] as a function of [Fe/H]. At [Fe/H] $>$ $-$2.5, scatters of [Zn/Fe] and [Mg/Fe] almost overlap with each other. Larger scatters of [Zn/Fe] than those of [Mg/Fe] at [Fe/H] $\lesssim$ $-$2.5 are caused by the ejection of Zn from ECSNe to the ISM that is still inhomogeneous in metallicity. This result suggests that scatters seen in \citet{2017A&A...606A..71S} are mostly caused by observational errors.

At [Fe/H] $\gtrsim$ $-$1, most stars have [Zn/Fe] $\sim$ $-$0.4 (Figure \ref{ZnFe}). This feature reflects the increase of the yield of Zn in CCSNe at higher metallicities (Figure \ref{Yields}). {This increase of Zn yield is caused by the neutron-capture process during He and C burning at higher metallicity \citep{2006ApJ...653.1145K}.}

\begin{figure}[htbp]
\epsscale{1.0}
\plotone{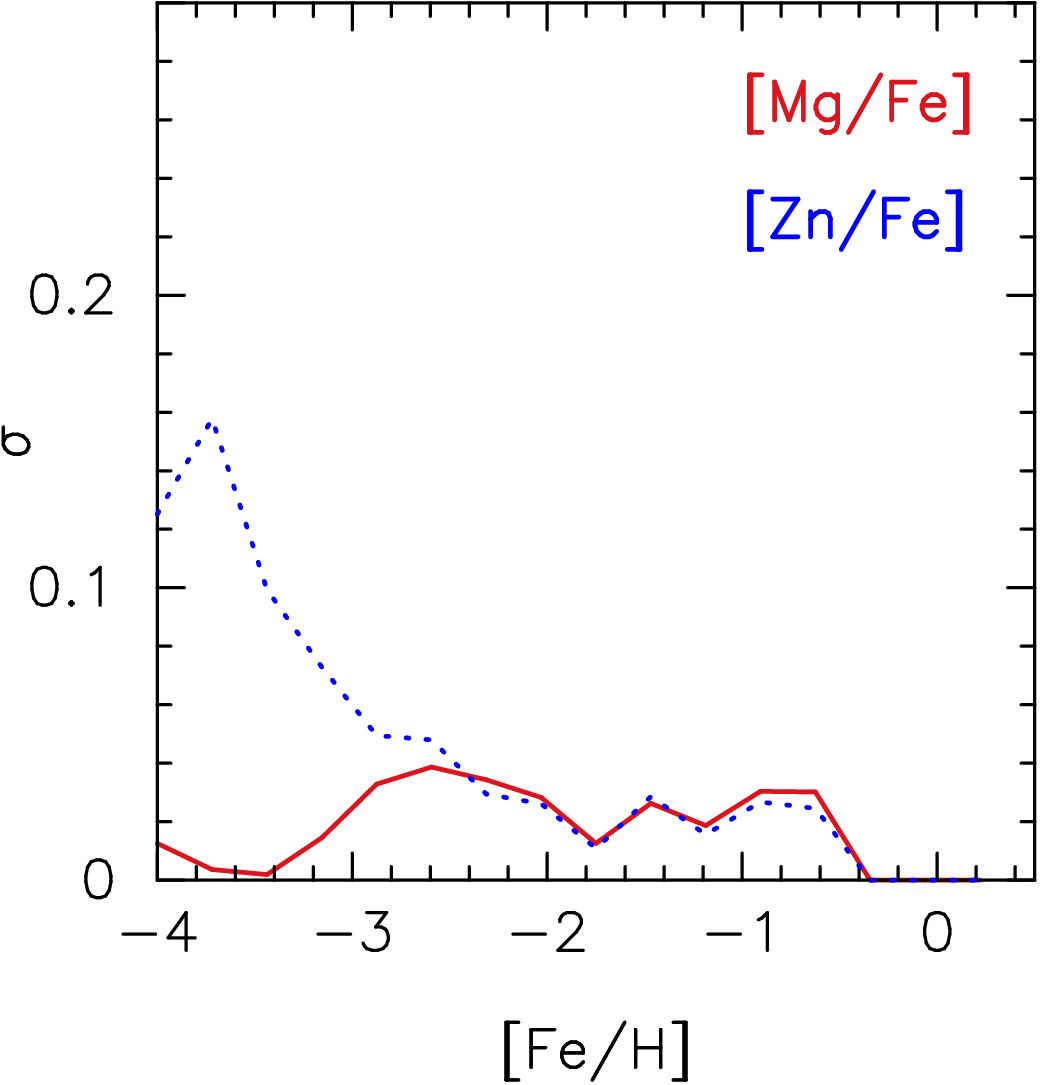}
\caption{{Standard deviations ($\sigma$) of [Mg/Fe] (the red solid curve) and [Zn/Fe] (the blue dotted curve) as a function of [Fe/H] in model A.}\label{Sigma}}
\end{figure}
\subsection{Astrophysical sites of Zinc}\label{site}
In this study, we consider ECSNe, HNe, and normal CCSNe as the astrophysical sites of Zn. As shown in Figure \ref{Yields}, ECSNe and HNe have higher ratios of [Zn/Fe] than those of normal CCSNe. This result means that ECSNe and HNe would have a significant impact on the enrichment of Zn. However, we do not know how these sites affect the enrichment of Zn in a galaxy. Rates of ECSNe and HNe are not well determined. This uncertainty would affect the enrichment history of Zn. Here we discuss how each expected source of Zn affects the enrichment of Zn in the galaxy.

Figure \ref{ZnFeHNe} shows [Zn/Fe] as a function of [Fe/H] for the models assuming that ECSNe do not contribute to the enrichment of Zn. Figure \ref{ZnFeHNe}a represents the result of model D. The HN fraction for model D ($f_{\rm HN}$ = 0.5) is taken from \citet{2006ApJ...653.1145K}. According to Figure \ref{ZnFeHNe}a, the [Zn/Fe] ratios are flat at low metallicities. The observed stars with [Zn/Fe] $\gtrsim$ 0.5 at [Fe/H] $\lesssim$ $-$2 cannot be explained, which is consistent with the result in \citet{2006ApJ...653.1145K}. Previous studies show that the increasing trend of the [Zn/Fe] ratios toward low metallicity can be explained if the yields of individual SNe are reflected in the abundances of EMP stars \citep{2002ApJ...565..385U, 2005ApJ...619..427U, 2007ApJ...660..516T}. However, our results suggest that the metal mixing erases these signatures even at [Fe/H] $\sim$ $-$3. 

\begin{figure}[htbp]
\epsscale{1.0}
\plotone{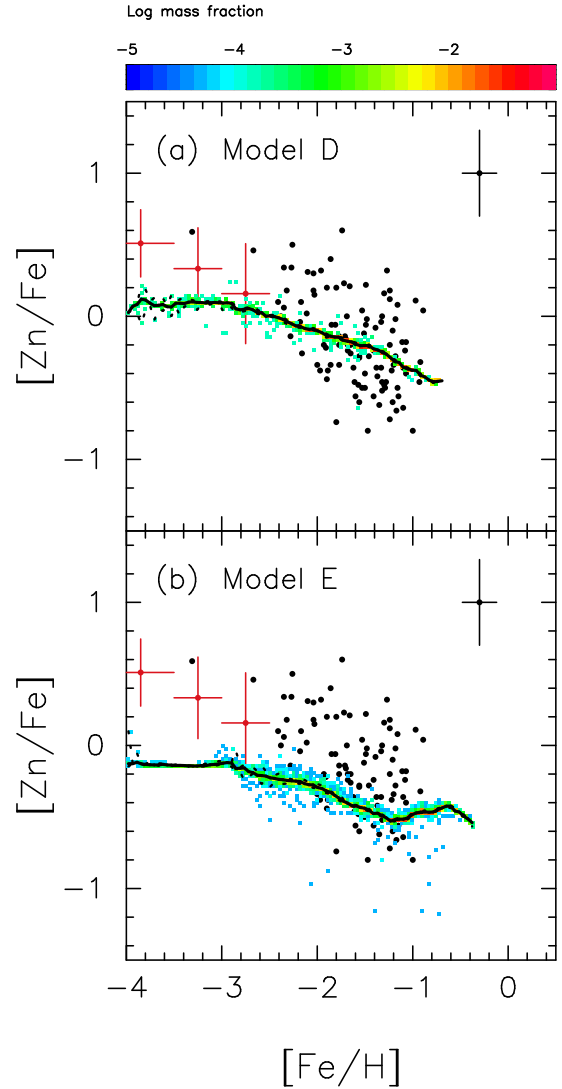}
\caption{Same as Figure \ref{ZnFe} but for the models without ECSNe. Panels (a) and (b) represent the models with $f_{\rm HN}$ = 0.5 and 0.05 (models D and E, respectively). \label{ZnFeHNe}}
\end{figure}

Figure \ref{ZnFeHNe}b shows the [Zn/Fe] ratios computed in model E. The HN fraction of model E ($f_{\rm HN}$ = 0.05) is taken to be consistent with the observed rate of long GRBs \citep{2004ApJ...607L..17P, 2007ApJ...657L..73G}. The [Zn/Fe] ratios of all stars at [Fe/H] $\lesssim$ $-$3 in this model are lower than those of the observation. This result suggests that if the rate of HNe is consistent with the observed rate of long GRBs, it is not possible to explain even the mean value of [Zn/Fe] in EMP stars without another source of Zn such as ECSNe.

We also consider another possibility that HNe do not contribute to the enrichment of Zn. This is because nucleosynthesis of Zn in HNe highly depends on the model parameters such as a position of mass cut. Moreover, HNe alone cannot explain the solar isotopic abundances of Zn because they cannot synthesize enough Zn isotopes except for $^{64}$Zn {at low metallicity \citep[e.g.,][]{2006ApJ...653.1145K,2011MNRAS.414.3231K} as discussed in Section \ref{Nucleosynthesis}.}

Figure \ref{ZnFenHNe} represents the [Zn/Fe] ratios computed by the models without HNe but with ECSNe. The values of [Zn/Fe] for [Fe/H] $<$ $-$2.5 in model F are lower than the observations. On the other hand, the values of [Zn/Fe] for [Fe/H] $<$ $-$2.5 in model G are high enough to be consistent with those in the observations. This difference is caused by the different rates of ECSNe in these models. The mass range of the progenitors of ECSNe in model F is 8.2 -- 8.4 $M_{\sun}$ at $Z$ = 0.0001. The fraction of ECSNe from this mass range corresponds to 6.9 \% of all CCSNe. Although the result is sensitive to the mass range of ECSNe in low metallicity, these results suggest that it is possible to explain the observed abundances of metal-poor stars without the contribution of HNe.

{Astrophysical sites of Zn may be more tightly constrained by examining the enrichment of Sr and other trans-iron elements with Zn. \citet{2018ApJ...852...40W} show that not only Zn but also other light trans-iron elements from Zn to Zr are enhanced in the ejecta of ECSNe. We find that the model A predicts the stars with [Fe/H] $<$ $-$3 having [Sr/Fe] $>$ $-$0.3. On the other hand, the observations of Local Group galaxies show that 65 \% of stars with [Zn/Fe] $>$ 0.5 have [Sr/Fe] $>$ $-$0.3 \citep[SAGA database,][]{2008PASJ...60.1159S, 2011MNRAS.412..843S, 2017PASJ...69...76S, 2013MNRAS.436.1362Y}. \citet{2017ApJ...837....8A} report that there is a diversity of the abundances of light neutron-capture elements from Sr to Pd. They show that the nucleosynthesis models with different electron fractions \citep{2011ApJ...726L..15W} or proto-neutron star masses \citep{2013ApJ...770L..22W} may explain such diversity. The $s$-process in fast-rotating massive stars may also cause a variation of [Sr/Fe] ratios \citep[e.g.,][]{2011Natur.472..454C, 2013A&A...553A..51C, 2014A&A...565A..51C}. We do not consider these possibilities in our simulations, which will be discussed in our forthcoming paper.}

\begin{figure}[htbp]
\epsscale{1.0}
\plotone{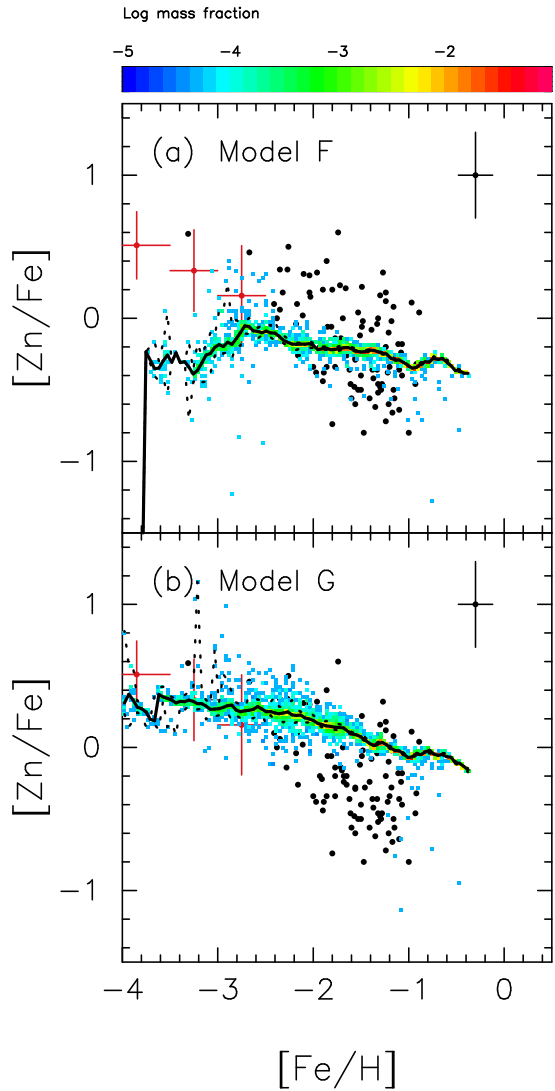}
\caption{Same as Figure \ref{ZnFe} but for the models without HNe. Panels (a) and (b) represent models F and G, respectively. \label{ZnFenHNe}}
\end{figure}

\subsection{Efficiency of metal mixing}\label{sec:mixing}
Determination of the efficiency of metal mixing is a long-standing issue in the study of galactic chemical evolution. \citet{2017ApJ...838L..23H} suggest that the value of $C_{\rm d}$ should be $C_{\rm d}$ $\gtrsim$ 0.01 to explain the observed $r$-process abundances in dwarf galaxies. This efficiency corresponds to the timescale of metal mixing of $\lesssim$ 40 Myr. Abundances of Zn in metal-poor stars will be another indicator to constrain the efficiency of metal mixing. Here we discuss the effect of metal mixing on the enrichment of Zn.

Figure \ref{ZnFemix} shows [Zn/Fe] versus [Fe/H] computed for the models with different efficiencies of metal mixing. Figure \ref{ZnFemix}a represents the result of model H. This model adopts a ten times lower value of diffusion coefficient of metal mixing ($C_{\rm d}$) than that of model A. Due to the lower efficiency of metal mixing, model H tends to have a larger fraction of stars with [Zn/Fe] $>$ 0.5 in EMP stars than other models do. Also, scatters of [Zn/Fe] for [Fe/H] $>$ $-$2 are more clearly seen in model H. \deleted{However, scatters of [Zn/Fe] seen in model H are larger than those of the observations.} Figure \ref{ZnFemix}b denotes the result of model I. {This model has smaller scatters of [Zn/Fe] in [Fe/H] $\lesssim$ $-$3 than those of models A and H.}\deleted{This model does not have high [Zn/Fe] stars in lower metallicity (Figure \ref{ZnFemix}b). This result implies that the efficiency of metal mixing is too high in model I.}

\begin{figure}[htbp]
\epsscale{1.0}
\plotone{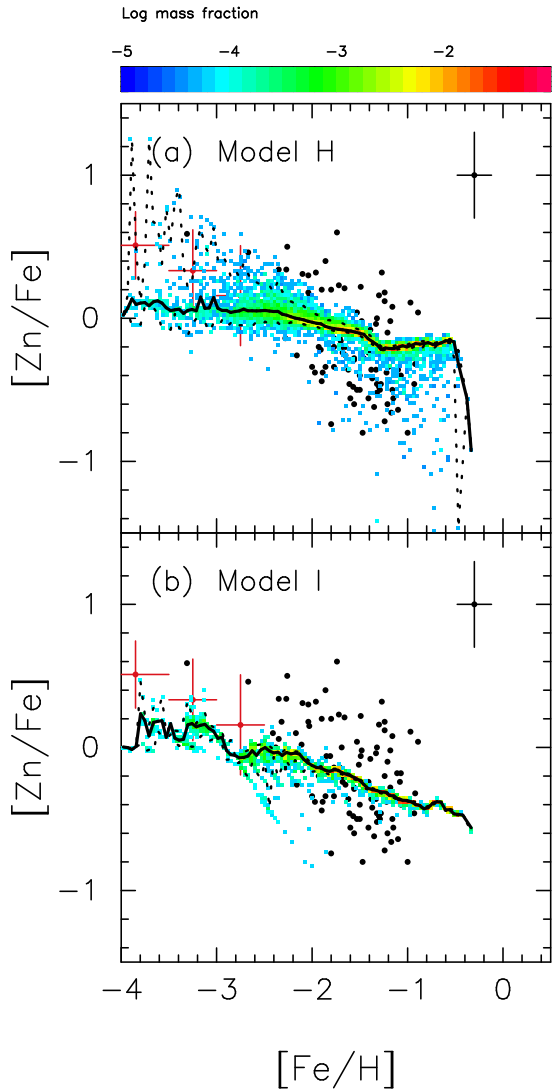}
\caption{Same as Figure \ref{ZnFe} but for the different efficiencies of metal mixing. Panels (a) and (b) show models H ($C_{\rm d}$ = 0.001) and I ($C_{\rm d}$ = 0.1), respectively. \label{ZnFemix}}
\end{figure}

According to these results, models with the value of {$C_{\rm d}$ $\sim$ 0.001 to 0.1 do not significantly deviate from the observed abundances of Zn. The value of $C_{\rm d}$ is consistent with the estimation from Ba abundances \citep{2017ApJ...838L..23H}, turbulent mixing layers, and turbulent channel flows \citep[e.g.,][]{1987JCoPh..71..343H}.} This result means that the value of $C_{\rm d}$ $\sim$ 0.01 appears to be suitable for SPH simulations of galaxies if we adopt the shear-based metal-diffusion model \citep{2010MNRAS.407.1581S, 2017AJ....153...85S}. \citet{2017ApJ...838L..23H} estimated that the timescale of metal mixing is $\approx$ 40 Myr for $C_{\rm d}$ $=$ 0.01. 

The value of $C_{\rm d}$ depends on the treatment of metal diffusion. We adopt the shear-based metal diffusion model \citep{2010MNRAS.407.1581S, 2017AJ....153...85S} in this study. On the other hand, \citet{2009MNRAS.392.1381G} constructed the model that used a velocity dispersion to estimate a diffusion coefficient. \citet{2016ApJ...822...91W} show that the diffusion coefficient estimated by a velocity dispersion based model is twice as large as that estimated by a shear-based model. They also show that the result is not strongly affected by the choice of a metal-diffusion model.

\section{Conclusions}\label{Conclusions}
We studied the enrichment histories of Zn in dwarf galaxies using a series of high-resolution chemodynamical simulations. We newly considered ECSNe as sources of Zn. The final stellar mass of our model was 5 $\times$ 10$^6$ $M_{\sun}$. This model is comparable to the observed natures (stellar masses, metallicity distributions, and $\alpha$-element evolutions) of Local Group dSphs such as the Sculptor and Leo I dSphs.

\deleted{We showed that models taking into account the contribution of ECSNe reproduced an increasing trend of [Zn/Fe] ratio toward lower metallicities (Figure \ref{ZnFe}). This trend is consistent with the observations.} We found that stars with [Zn/Fe] $\gtrsim$ 0.5 in our models reflected the nucleosynthetic abundances of ECSNe (Figure \ref{ZnFe}). In the early phase of galaxy evolution, gases with high [Zn/Fe] ratios owing to ECSNe remained due to the inhomogeneity of spatial distribution of metallicity.

Our results suggest that scatters of [Zn/Fe] in higher metallicities come from the contribution of SNe Ia. These stars were formed at $\lesssim$ 4 Gyr from the beginning of the simulation (Figure \ref{AgeFe}). {We found that the scatters of [Zn/Fe] are consistent with the scatters of [Mg/Fe] at [Fe/H] $\gtrsim$ $-$2.5.}

In this study, we examined the contribution from several astrophysical sources of Zn. If we do not take into account the contribution of ECSNe, we cannot reproduce the increasing trend of [Zn/Fe] toward lower metallicities (Figure \ref{ZnFeHNe}). On the other hand, the observed trend of [Zn/Fe] could be reproduced without assuming the production of Zn from HNe (Figure \ref{ZnFenHNe}). These results suggest that ECSNe can be one of the contributors to the enrichment of Zn in galaxies.

We also studied the efficiencies of metal mixing in galaxies. Our result suggests that the scaling factor for metal diffusion ($C_{\rm d}$) should be $\approx$ 0.01 to explain the presence of stars with [Zn/Fe] $\gtrsim$ 0.5 (Figure \ref{ZnFemix}). This efficiency corresponds to the timescale of metal mixing of $\approx$ 40 Myr \citep{2017ApJ...838L..23H}.

\twocolumngrid
\acknowledgements
We are grateful for the referee who gives insightful comments to improve the manuscript. This work was supported by JSPS KAKENHI Grant Numbers 15J00548, 26707007, 17K05391, 26400237, MEXT SPIRE, JICFuS, and the RIKEN iTHES project. This study was also supported by JSPS and CNRS under the Japan-France Research Cooperative Program. Numerical computations and analysis were in part carried out on Cray XC30 and computers at Center for Computational Astrophysics, National Astronomical Observatory of Japan. This research has made use of NASA's Astrophysics Data System.
\appendix
\twocolumngrid
Here we show the effects of different SN yields of Zn (Appendix \ref{sec:yields}) and dependence on the spatial resolution of simulation (Appendix \ref{sec:resolution}). Table \ref{appendix} lists models discussed in the Appendix.

\begin{deluxetable}{llrl}[htbp]
   \tabletypesize{\scriptsize}
   \tablecaption{List of models discussed in the Appendix. \label{appendix}} 
   \tablecolumns{2}
   \tablewidth{0pt}
   \tablehead{
      \colhead{Model}&
      \colhead{$N$}&
      \colhead{$\epsilon_{\rm g}$}& 
      \colhead{SN Yields}\\
      \colhead{}&\colhead{}&\colhead{pc}&\colhead{}}
      \startdata
      A\tablenotemark{*}&2$^{18}$&7.8&N13\\ 
      J&2$^{18}$&7.8&CL04\\                               
      K&2$^{17}$&9.9&N13\\
      L&2$^{16}$&14.1&N13\\
      \enddata
      \tablenotetext{*}{Model A is the same model as listed in Table \ref{models}.}
     \tablecomments{From left to right, columns show names of models, initial total number of particles, gravitational softening length, and adopted SN yields.}

\end{deluxetable}
\section{Effects of supernova yields of Zinc}\label{sec:yields}
As we have shown in Figure \ref{Yields}, SN yields of Zn and Fe depend on stellar evolution models. This can affect the results of this study. Figure \ref{ZnFeSNe} compares the computed evolution of [Zn/Fe] versus [Fe/H] in models A and J. Models A and J adopt SN yields of N13 and CL04, respectively. As shown in Figure \ref{ZnFeSNe}, both models have the increasing trend toward lower metallicities. Model J has a $\sim$ 0.3 dex lower median value of [Zn/Fe] than that of model A. In addition, model J has larger scatters in [Zn/Fe] at [Fe/H] $>$ $-$0.5 than those in model A. These differences reflect the lower production of Zn in CL04 than that in N13. However, both results are within the range of scatters in observed [Zn/Fe]. This result implies that the effects of the difference of SN yields do not substantially affect the enrichment history of Zn.
\begin{figure}[htbp]
\epsscale{1.0}
\plotone{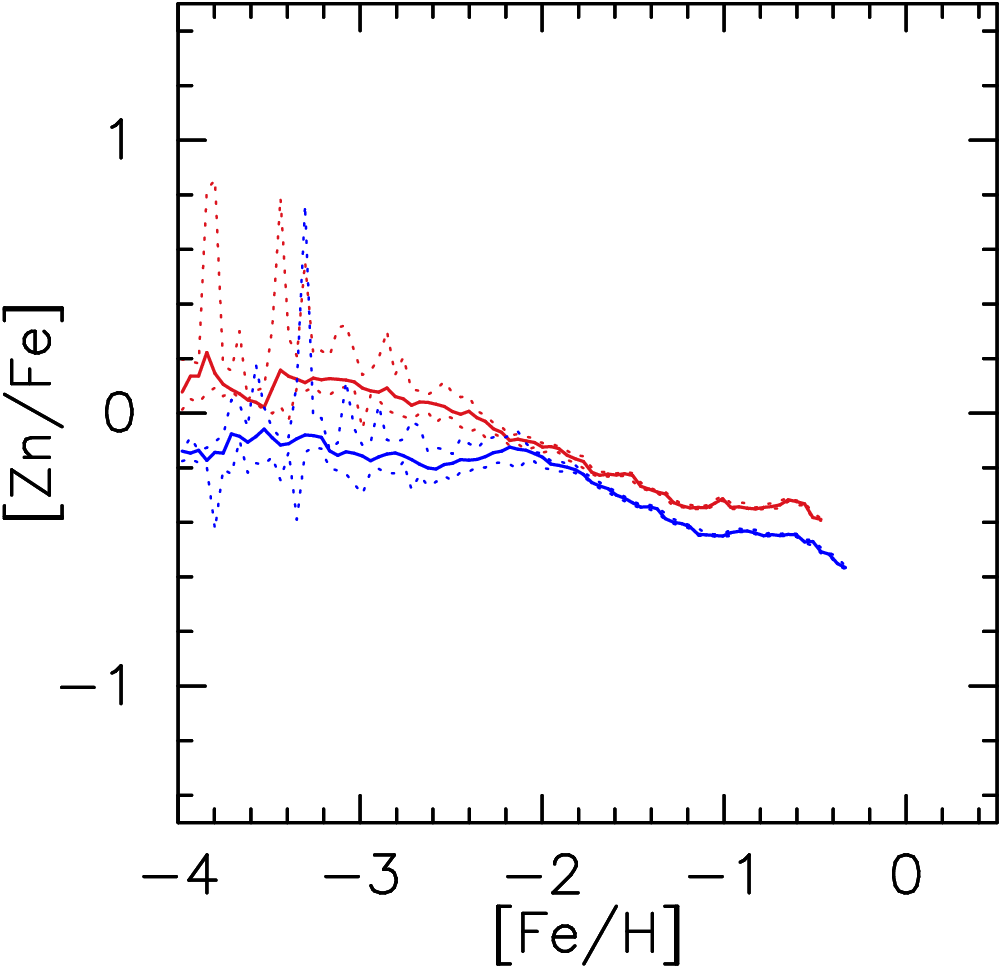}
\caption{Comparison of the [Zn/Fe] evolution for the models assuming different SN yields. Model A (red curves) adopts the nucleosynthesis yield of N13. Model J (blue curves) adopts the yield of CL04. Solid curves show the median values. Dotted curves indicate the 5 \% and 95 \% significance levels at each [Fe/H] bin.\label{ZnFeSNe}}
\end{figure}
\section{Dependence on the resolution}\label{sec:resolution}
Figure \ref{ZnFeresolution} compares the [Zn/Fe] evolutions for different resolution models. The levels of scatters in [Zn/Fe] do not substantially differ among these models. The number fraction of stars with [Fe/H] $\gtrsim$ $-$0.6 seen in model K is less than 0.04, i.e., the number of these stars is negligible. We therefore conclude that the spatial resolution of simulations does not change our main results.
\begin{figure}[htbp]
\epsscale{1.0}
\plotone{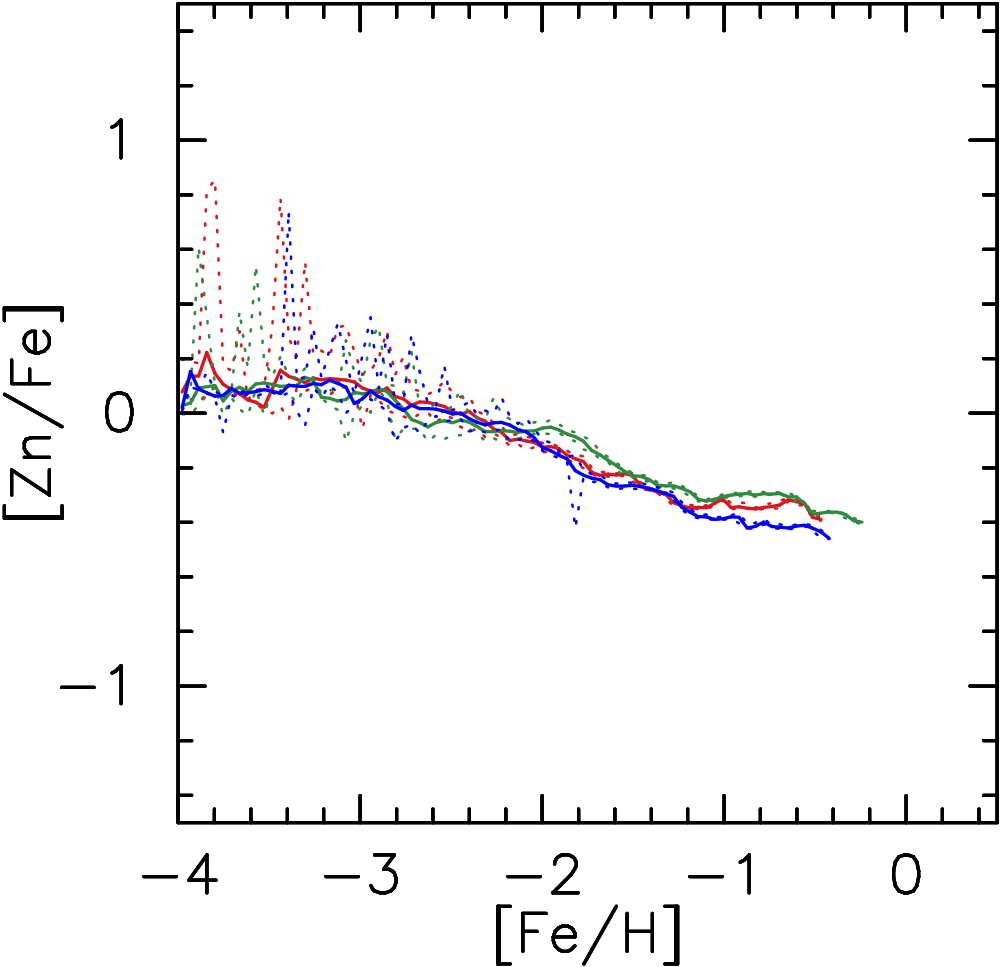}
\caption{Same as Figure \ref{ZnFeSNe} but for models with different resolution. Red, green, and blue curves represent models A ($N$ = $2^{18}$), K ($N$ = $2^{17}$), and L ($N$ = $2^{16}$), respectively. \label{ZnFeresolution}}
\end{figure}
\mbox{}\newpage

\bibliography{sampleNotes}

\listofchanges

\end{document}